\documentclass[twocolumn,preprintnumbers,amsmath,amssymb,nofootinbib]{revtex4}

\usepackage{graphicx}
\usepackage{dcolumn}
\usepackage{bm}
\usepackage{mathrsfs}

\begin{document}


\title{Seesaw mechanism at electron-electron colliders}

\author{Takehiko Asaka}
\email{asaka@muse.sc.niigata-u.ac.jp}
\affiliation{Department of Physics, Niigata University, 950-2181 Niigata, Japan}

\author{Takanao Tsuyuki}%
 \email{tsuyuki@muse.sc.niigata-u.ac.jp}
 \affiliation{Graduate School of Science and Technology, Niigata University, 950-2181 Niigata, Japan}

\date{August 20, 2015}

\begin{abstract}
  We consider the Standard Model with right-handed neutrinos  to explain the masses of active neutrinos by the  seesaw mechanism.  Since active neutrinos as well as heavy neutral  leptons are Majorana fermions in this case, the lepton number violating process can be induced. We discuss the inverse neutrinoless double beta decay  $e^- e^- \to W^- W^-$ in the framework of the seesaw mechanism and  its detectability at future colliders.  It is shown that the cross  section can be 17 fb for $\sqrt{s}=3$ TeV even with the  stringent constraint from the neutrinoless double beta decays if  three (or more) right-handed neutrinos exist.  In such a case, the  future $e^- e^-$ colliders can test lepton number violation mediated  by a right-handed neutrino lighter than about 10~TeV.
\end{abstract}


\maketitle

\section{Introduction}
The origin of neutrino masses is one of the most important questions
in particle physics at present. Various oscillation experiments have provided the mass squared differences and mixing angles of active
neutrinos very precisely \cite{Gonzalez-Garcia:2014bfa}; however, the
mass ordering, violation of $CP$ symmetry and fundamental property
of massive neutrinos ({\it i.e.}, Majorana or Dirac particles) are still
unknown. The simplest way to explain the neutrino masses is to add
right-handed neutrinos to the Standard Model (SM). The smallness of
active neutrino masses can be explained by the large Majorana masses of
right-handed neutrinos thanks to the seesaw mechanism
\cite{Minkowski:1977sc,Yanagida:1979as,Yanagida:1980xy,GellMann:1980vs,Ramond:1979,Glashow:1979nm,Mohapatra:1979ia}.

In the seesaw mechanism, the mass eigenstates of neutrinos are three active neutrinos which have tiny masses observed in oscillation experiments, and heavy neutral leptons (HNLs) which are almost identical to the right-handed states. The HNL masses are determined by Majorana masses and independent from the electroweak Higgs mechanism. They can take arbitrary values as long as
they are sufficiently heavy to realize the seesaw mechanism. This is
the reason why such HNLs can cause interesting phenomena in various
aspects of particle physics and cosmology.

Right-handed neutrinos can explain the baryon asymmetry of the
Universe.  The well-known scenario is the canonical
leptogenesis~\cite{Fukugita:1986hr} in which they need to be
heavier than ${\cal O}(10^9)$ GeV~\cite{Giudice:2003jh} (or ${\cal O}(10^6)$~GeV when the
nonthermal production is realized~\cite{Leptogenesis_inflation_decay}).  The resonant
leptogenesis with quasidegenerate right-handed
neutrinos~\cite{Pilaftsis:2003gt} can be effective with much smaller
masses.  Moreover, if we use the flavor oscillation of right-handed
neutrinos, the required mass can be as small as ${\cal O}(1)$ MeV~
\cite{Akhmedov:1998qx,Asaka:2005pn,Asaka:2013jfa}.

In addition, a right-handed neutrino can play
 an important role in
astrophysics.  It can be the dark matter candidate with $
{\cal O}(1)$~keV mass~\cite{Dodelson:1993je}.  This particle can also explain other phenomena, such as the pulsar kick~\cite{Pulsarkick} (for a review, see Ref.~\cite{Kusenko:2009up}). Further, right-handed neutrinos with ${\cal O}(0.1)$ GeV
may be important for the supernova explosion~\cite{Fuller:2009zz}.

Right-handed neutrinos are required by grand unified theories based on SO(10) \cite{Georgi,Fritzsch:1974nn} or larger symmetry groups. These particles can also accomplish the
bottom-tau Yukawa unification \cite{Tsuyuki:2014xja}
which is necessary for realistic grand unified theories.
It has been discussed that they are favored to be lighter than ${\cal O}(10^7)$ GeV
due to the argument of the electroweak naturalness~\cite{Clarke:2015gwa}.

In this way, right-handed neutrinos in the seesaw mechanism are
well-motivated particles beyond the SM.  The direct test
of HNLs is possible if they are light and their mixings are
large.  So far various experiments have been performed to search
for HNLs (for example, see reviews~\cite{Kusenko:2004qc,Gorbunov:2007ak,Atre:2009rg,Deppisch:2015qwa}).

The noble consequence of the seesaw mechanism is that active neutrinos
and HNLs are Majorana particles, and hence the lepton number violating processes can appear. A 
well-known example is the neutrinoless double beta ($0\nu\beta\beta$)
decay $(Z,A) \to (Z+2,A) + 2 e^-$, which violates the lepton number by
two units~\cite{Pas:2015eia}.

\begin{figure}[tb]
\includegraphics[width=8.6cm]{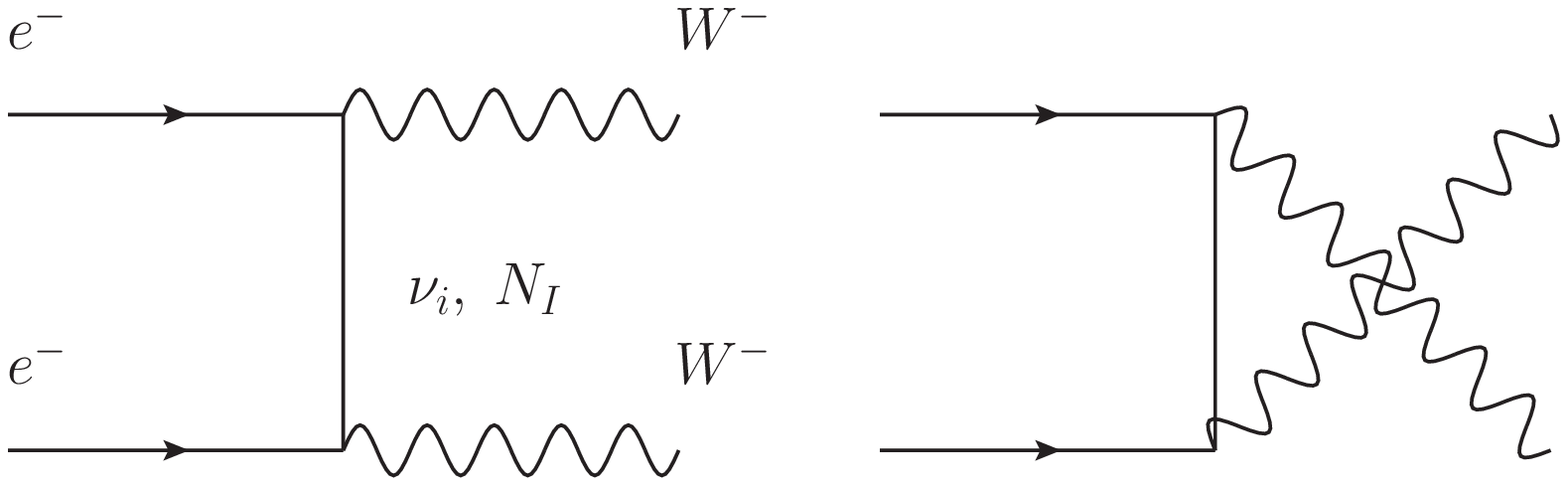}
\caption{\label{FeeWW} 
  Feynman diagrams for $e^-e^-\to W^-W^-$.
}
\end{figure}

In this paper, we consider another example of such processes,
\begin{align}
e^-e^-\to W^- W^-, \label{eeeww}
\end{align}
which is called the ``inverse neutrinoless double beta decay''
\cite{Rizzo:1982kn} (see Fig. \ref{FeeWW}), which cannot also happen in the SM. The $e^- e^-$ collision at high
energy is one attractive option of the future lepton colliders
such as the International Linear
Collider (ILC \cite{Baer:2013cma}) and the Compact Linear Collider
(CLIC \cite{Accomando:2004sz}).  Various aspects of the process
(\ref{eeeww}) have been studied so
far~\cite{Rizzo:1982kn,London:1987nz,Dicus:1991fk,Belanger:1995nh,Greub:1996ct,Gluza:1995ky,Gluza:1995ix,Gluza:1995js,Rodejohann:2010jh,Banerjee:2015gca}.

The advantages of the process (\ref{eeeww}) over the $0\nu\beta\beta$
decay for the test of the lepton number violation are (i) the signal
event is clean and (ii) the prediction is free from the
uncertainty in the nuclear matrix element.  Furthermore, we should
mention that these processes are complementary tests for the lepton
number violation.  As we will show, HNLs can give a
 destructive contribution to the $0\nu\beta\beta$ decay so
that no signal could be observed.  Even in this
case, the inverse $0\nu\beta\beta$ decay could be observed since the energy
scales of these processes are different.

In this paper, we shall revisit the inverse $0\nu\beta\beta$ decay paying
special attention to the following points: (i) We study the process in
the concrete model, {\it i.e.}, the SM extended by ${\cal N}$
right-handed neutrinos with the seesaw mechanism, and try to derive
the theoretical prediction being specific to the model.  (ii) We take
fully into account the interference effects between active neutrinos
and HNLs for both the $0\nu\beta\beta$ and the inverse $0\nu\beta\beta$
processes. (iii) We estimate the upper bounds on the cross section depending on the number of right-handed neutrinos. We use corrected cross section (for the difference from previous studies, see the Appendix).  In addition we take into account the possibilities of the
fine-tuning in parameters of the model, which have not been studied
thoroughly. 

We find that the maximal cross
section can be 0.47 fb (17 fb) for the center-of-mass energy
$\sqrt{s}=$ 500 GeV (3 TeV) avoiding the stringent constraint from the
$0\nu\beta\beta$ decay if there are three right-handed neutrinos. This
is a contrast to the previous results.  Therefore, the process
(\ref{eeeww}) can be a good target of the future lepton colliders.

The rest of this paper is organized as follows.
In Sec. \ref{sSM}, we define the model parameters and list existing
constraints on them. In Sec \ref{sInv}, we derive upper bounds on the
cross sections of the process (\ref{eeeww}) in the cases with one,
two, three right-handed neutrinos in turn. 
Finally, Sec. \ref{sCon} is devoted to conclusions.
We add an Appendix to present the definitions of variables 
in the cross section of $e^- e^- \to W^- W^-$.

\section{Standard Model with Right-handed Neutrinos} \label{sSM}
Let us first explain the framework of the present analysis.
We consider the SM extended by 
right-handed neutrinos $\nu_R$ with Lagrangian
\begin{eqnarray}
  \label{eq:LAG}
  {\cal L} &=& {\cal L}_{\rm SM}
  + 
  \overline \nu_{R I} i \partial_\mu \gamma^\mu 
  \nu_{R I}
  \nonumber \\
  &&
  - F_{\alpha I} \overline L_\alpha \Phi \nu_{R I}
  - \frac{M_I}{2} \overline \nu_{R I} \nu_{R I}^c
  + h.c. \,.
\end{eqnarray}
Here we shall assume $|M_{D\alpha I}| \equiv |F_{\alpha I}| \langle
\Phi \rangle \ll M_I\; (\alpha=e,\mu,\tau;\; I=1, \cdots, {\cal N})$
in order to realize the seesaw mechanism.  In this case the mass
eigenstates of neutrinos are three active neutrinos $\nu_i$ ($i =
1,2,3$) with masses $m_i$ and ${\cal N}$ HNLs
$N_I$ with masses $\simeq M_I$. 
The mass ordering of HNLs can be chosen as
$M_1 \le M_2 \le  \cdots \le M_{\cal N}$ without loss of generality.
The mixing of neutrinos in the charged current interactions is then
written as
\begin{eqnarray}
  \nu_{L \alpha }
  = \sum_{i=1}^3 U_{\alpha i} \nu_i 
  +
  \sum_{I=1}^{\cal N}
  \Theta_{\alpha I} N_I^c \,,
\end{eqnarray}
where $U$ is the Pontecorvo-Maki-Nakagawa-Sakata (PMNS) mixing matrix of active neutrinos,
while $\Theta_{\alpha I}$ represent the mixings of HNLs obtained by the diagonalization of the $(3+{\cal N})\times (3+{\cal N})$ neutrino mass matrix including radiative corrections.

In this model, the masses and mixings of active neutrinos and HNLs
must satisfy the following relations:
\begin{align} 
  \label{eq:seesaw} 
  0=\sum_i^3 U_{\alpha i}U_{\beta
    i}m_i+\sum_{I=1}^\mathcal{N}\Theta_{\alpha I} \Theta_{\beta I}M_I
  \,.
\end{align}
These relations hold if the Majorana mass terms for the left-handed
neutrinos are absent at the tree level in the canonical seesaw mechanism discussed
here.  We should note that such Majorana masses are induced by radiative corrections  \cite{Grimus:2002nk,AristizabalSierra:2011mn} and would alter the relation (\ref{eq:seesaw}) desperately.  We assume that such corrections are sufficiently suppressed by artificial fine-tuning. 

For the inverse $0\nu\beta\beta$ decay (\ref{eeeww}), only the component
with $\alpha = \beta = e$,
\begin{align} 
  \label{Eseesaw2}
  0 = m_{\rm eff}^\nu+\sum_I \Theta_{eI}^2 M_I \,,
\end{align}
is relevant. We call this equation ``the
seesaw relation'' from now on. We define $m_{\rm eff}^\nu$ by 
\begin{eqnarray}
  m_{\rm eff}^\nu \equiv \sum_{i=1}^3 U_{ei}^2 \, m_i 
  \label{Emeff} \,.
\end{eqnarray}

In the considered model both active neutrinos
and HNLs are Majorana fermions, and then they induce $0 \nu\beta\beta$ decay.
The half-life of the decay is expressed as~%
\begin{eqnarray}
  T_{1/2}^{-1} = 
  {\cal A} \frac{m_p^2}{\langle p^2 \rangle^2}
  | m_{\rm eff} |^2 \,.
\end{eqnarray}
Here and hereafter, we use the notation and the results
for the $0\nu\beta\beta$ decay given in Ref.~\cite{Faessler:2014kka}.
The effective neutrino mass in the considered model
is given by
\begin{eqnarray}
  \label{eq:meff}
  m_{\rm eff} = m_{\rm eff}^\nu + 
  \sum_{I=1}^{\cal N} \Theta_{eI}^2 \, M_I \, f_I \,,
\end{eqnarray}
Here the first term represents the contribution from the active
neutrinos which is given by Eq.~(\ref{Emeff}), and the second term 
denotes the contributions from HNLs, in which the
suppression of the nuclear matrix element 
for $M_I \gg \sqrt{\langle
  p^2 \rangle} \simeq 200$ MeV is taken into account
by the function $f_I$~\cite{Faessler:2014kka},
\begin{eqnarray}
  f_I \simeq \frac{\langle p^2 \rangle}{\langle p^2 \rangle + M_I^2}\,.
\end{eqnarray}

As for the $0 \nu\beta\beta$ decay of $^{136}$Xe, the lower bound on the
half-life is $T_{1/2} > 3.4 \times 10^{25}$ yr at 90\%
C.L.~\cite{Gando:2012zm}, which can be translated into the upper bound
on the effective neutrino mass as
\begin{eqnarray}
  \label{eq:meffUB}
  |m_{\rm eff}| < m_{\rm eff}^{\rm UB} = 
  (0.185-0.276 )~\mbox{eV} \,,
\end{eqnarray}
where we have taken into account the uncertainties
estimated in~\cite{Faessler:2014kka}.
On the other hand, the half-life bound of $^{76}$Ge is
$T_{1/2} > 3.0 \times 10^{25}$ yr at 90\% C.L.~\cite{Agostini:2013mzu}.
In this case the bound on $|m_{\rm eff}|$ is 
\begin{eqnarray}
  |m_{\rm eff}| < m_{\rm eff}^{\rm UB} = 
  (0.213-0.308 )~\mbox{eV} \,.
\end{eqnarray}
We then apply the bound (\ref{eq:meffUB}) throughout this analysis 
in order to make the conservative analysis.

Furthermore, the mixings of HNLs are constrained by 
various 
experiments~\cite{Kusenko:2004qc,Gorbunov:2007ak,Atre:2009rg,Deppisch:2015qwa}.
In the following analysis, we consider the case when the mass of the lightest HNL
is $M_1 \gtrsim 3$ GeV in order to avoid the stringent
constraints from the search experiments by using the decays of $\pi$,
$K$ and $D$ mesons.  Even in this case, there is an upper limit on
$|\Theta_{eI}|^2$ in the region $M_I<m_Z$ from the search for HNLs at
the Large Electron-Positron collider (LEP) experiment by the decay of the $Z$ boson~\cite{Abreu:1996pa}.  In the mass
range $M_I \simeq $ 6-50~GeV, the bound is given as $|\Theta_{eI}|^2 <
2.1\times 10^{-5}$ at 95\% C.L.  There is also an upper bound on the
mixing angle which comes from the electroweak precision tests. The
recent bound at 90~\% C.L. ~\cite{Antusch:2014woa} is
\begin{eqnarray}
  \label{eq:THe_EW}
  |\Theta_e|^2\equiv\sum_I |\Theta_{eI}|^2 
  < |\Theta_e|^2_{\rm EW}=2.1 \times 10^{-3} \,.
\end{eqnarray}
These are the upper bounds on the mixings
which are important for the analysis below.\footnote{
Our analysis can also be applied to processes $\mu^-\mu^-\to W^-W^-$
and $e^-\mu^-\to W^-W^-$ as in Ref. \cite{Rodejohann:2010jh}. The
latter process is, however, strongly constrained by the $\mu\to
e\gamma$ experiment \cite{Antusch:2014woa,Adam:2013mnn}, $|\sum_I \Theta_{e I}^*\Theta_{\mu I}|<10^{-5}$ (90\% C.L.).}

\section{Inverse Neutrinoless Double Beta Decay $e^- e^- \to W^-
  W^-$} \label{sInv} 
Now, we are in the position to discuss the inverse neutrinoless double
beta decay (\ref{eeeww}).  The cross section of the process in the
model with ${\cal N}$ right-handed neutrinos is given by
\begin{eqnarray}
  \label{Edsig}
  \frac{d \sigma_{\cal N}}{d \cos \theta}
  &=& 
  \frac{G_F^2 \, \beta_W}{32 \pi}
  \Bigl[ \,
  | A_t |^2 \, B_t + |A_u |^2 \, B_u 
  \nonumber \\
  &&~~~~~~~~~~~
  + ( A_t \, A_u^\ast + h.c. ) \, B_{tu} \Bigr] \,.
\end{eqnarray}
Here the definitions of variables in this equation are presented in the
Appendix.  Notice that only $A_t$ and $A_u$ depend on the parameters
of HNLs, namely, $M_I$ and $\Theta_{eI}$.

In the following analysis, 
we consider the detectability of $e^- e^- \to W^- W^-$
by taking the center-of-mass energy as
$\sqrt{s} = 0.5$, 1 and 3~TeV.
We assume the integrated luminosity of~100 fb, which can be achieved by
${\cal O}(10^7)$~s run of the ILC or CLIC~\cite{Agashe:2014kda} (the luminosities of the $e^- e^-$ colliders are expected to be the same order as $e^+ e^-$ colliders~\cite{Accomando:2004sz}).

Before discussing the realistic cases, we shall consider the two
extreme cases.  The first one is the limit when all HNLs are
sufficiently heavy ({\it i.e.}, $M_I \gg \sqrt{s}$) and only three
active neutrinos effectively take part in the process.  In this case
the cross section is given by~\cite{Belanger:1995nh}%
\footnote{ The prefactor of $\sigma_0$ in Eq.~(\ref{eq:sig_0}) is
  three times larger than the estimations
  in Refs.~\cite{Belanger:1995nh,Rodejohann:2010jh} (see the discussion in the Appendix).
}
\begin{eqnarray}
  \label{eq:sig_0}
  \sigma_0 = \frac{3 \, G_F^2 \, |m_{\rm eff}^\nu|^2}{4 \pi}
  =
  0.96 \times 10^{-18} \, \mbox{fb} \,
  \left( \frac{|m_{\rm eff}^\nu|}{0.28 \, \mbox{eV}} \right)^2 \,.
\end{eqnarray}
The cross section is determined solely by the effective
neutrino mass $m_{\rm eff}^\nu$ of active neutrinos in the $0 \nu\beta\beta$ decay.
It is seen that $\sigma_0$ is too small to be accessible
by future colliders.

The other limit is that all HNLs are sufficiently light
({\it i.e.}, $M_I \ll m_W \ll \sqrt{s}$).  
In this case it is found from the seesaw relation (\ref{Eseesaw2}) that
\begin{eqnarray}
  A_t = \frac{1}{t} 
  \left( m_{\rm eff}^\nu + \sum_{I=1}^{\cal N} \Theta_{eI}^2 \, M_I 
    \right)= 0 
\end{eqnarray}
and then $A_u = 0$, which results in the vanishing cross section.  In
other words, the seesaw relation (\ref{Eseesaw2}) ensures the
unitarity of the process (\ref{eeeww}) at high energies $\sqrt{s} \gg
M_I$~\cite{Belanger:1995nh}.

From now on, we will study the cross section
of $e^- e^- \to W^- W^-$ in the framework of the seesaw model
with ${\cal N} =$ 1, 2, and 3 right-handed neutrino(s).
Especially, we shall discuss the impacts 
of various constraints on the masses and mixings of HNLs 
discussed in the previous section.

\subsection{Case with one right-handed neutrino}
Let us first consider the SM with only one right-handed neutrino.  In
this case, there is one HNL $N_1$ with mass $M_1$ and mixing
$\Theta_{\alpha 1}$ in addition to active neutrinos.  This model
is incompatible with the oscillation data since there appear two
massless states among active neutrinos. We shall, however, discuss it 
in order to make our point clear.

\begin{figure}
\includegraphics[scale=1.4]{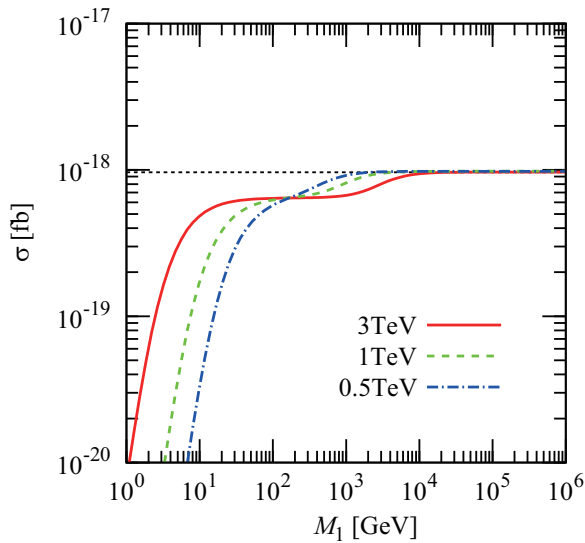}
\caption{\label{fig:SIG_1RHN_wSS} 
Cross section of $e^- e^- \to W^- W^-$ in the model with 
one right-handed neutrino (${\cal N}=1$) for $\sqrt{s} = 3$ TeV (the red solid line),
1 TeV (the green dashed line), and 0.5 TeV (the blue dotted-dashed line).
The horizontal line (the black dotted line) represents 
$\sigma_0$ in (\ref{eq:sig_0}).
Here we take $|m_{\rm eff}^\nu| = 0.276$ eV.
}
\end{figure}

One might think that $N_1$ with a relatively large mixing 
allowed by Eq. (\ref{eq:THe_EW}) may give rise to a significant contribution
to $e^- e^- \to W^- W^-$.  This is, however, not true in the
considered case.  The seesaw relation (\ref{Eseesaw2}) tells that
the mixing cannot be large but $\Theta_{e1}^2 = - m_{\rm eff}^\nu /M_1$,
and then $A_t$ is suppressed as
\begin{eqnarray}
  \label{eq:1RHN_at}
  A_t  = m_{\rm eff}^\nu 
  \left( \frac{1}{t} - \frac{1}{t - M_1^2}  \right)\,.
\end{eqnarray}
This shows that the cross section becomes independent on 
the mixing of $N_1$ but is determined from the effective mass
$m_{\rm eff}^\nu$ and $M_1$.
In Fig.~\ref{fig:SIG_1RHN_wSS}  we show the cross section
$e^- e^- \to W^- W^-$ with $|m_{\rm eff}^\nu|=0.276$ eV. 

It is seen that the cross section approaches to 
$\sigma_0$ (\ref{eq:sig_0}) for $M_1 \gg \sqrt{s}$ due to the decoupling
of $N_1$ as mentioned before.
On the other hand, when $M_1 \ll m_W $, the cross section 
behaves as
\begin{eqnarray}
  \sigma_1 = \frac{1}{6 \pi} 
  \frac{G_F^2 \, |m_{\rm eff}^\nu|^2 \, M_1^4 \, s^2}{m_W^8} \,,
\end{eqnarray}
and it vanishes for $M_1 \to 0$ as expected. 

In the model with ${\cal N}=1$, therefore, 
the cross section is too small to be observed at future colliders.
This is the direct consequence of the seesaw mechanism, namely, 
the mixing $\Theta_{e1}$ is determined by $m_{\rm eff}^\nu$
from the seesaw relation (\ref{Eseesaw2}).

\subsection{Case with two right-handed neutrinos}
Next, we discuss the case with two right-handed neutrinos, where there
appear two HNLs $N_1$ and $N_2$.  Notice that the lightest active neutrino in this
case is massless ({\it i.e.}, $m_1 = 0$ for the normal hierarchy and
$m_3 = 0$ for the inverted hierarchy).

In this case, by using the seesaw relation (\ref{Eseesaw2}) and
$|\Theta_e|^2$ (\ref{eq:THe_EW}), 
the mixing angles of $N_1$ and $N_2$ are given by%
\footnote{
$\Theta_{e1}^2$ can be taken to be real and positive without 
loss of generality.  Here we consider the case when
$\Theta_{e2}^2$ is real and negative.
The importance of the relative phases between 
the mixings has been discussed in Refs.~\cite{Gluza:1995ix,Gluza:1995js}.}
\begin{eqnarray}
  \Theta_{e1}^2
  &=&
  + \frac{M_2}{M_1+M_2} \, |\Theta_e|^2
  - \frac{m_{\rm eff}^\nu}{M_1+M_2} \,,
  \\
  \Theta_{e2}^2
  &=&
  - \frac{M_1}{M_1+M_2} \, |\Theta_e|^2
  - \frac{m_{\rm eff}^\nu}{M_1+M_2} \,.
\end{eqnarray}
We find that, when $M_1 \ll M_2$ and $|m_{\rm eff}^\nu|/M_1 \ll
|\Theta_e|^2$, the large mixing of $N_1$, $\Theta_{e1}^2 \simeq
|\Theta_e|^2_{\rm EW}$, can be realized.
This is because of the cancellation between the contributions of $N_1$
and $N_2$ in the seesaw relation
(\ref{Eseesaw2})~\cite{Dicus:1991fk,Belanger:1995nh}.  Some mechanism is desirable to stabilize such a fine-tuning, for example, by a discrete flavor symmetry.  This
issue is, however, beyond the scope of our analysis (similar situations have been discussed in Refs. \cite{Gluza:2002vs,Kersten:2007vk}).

The function $A_t$ is now
\begin{eqnarray}
  \label{eq:2RHN_At}
  A_t &=& - \frac{(M_2 - M_1) \, M_1 \, M_2}{(t-M_2^2)(t-M_1^2)} \,
  |\Theta_e|^2
  \\
  &&
  - \frac{ m_{\rm eff}^\nu [ (M_2^2 - M_1 \, M_2 +M_1^2) \, t
    - M_1^2 \, M_2^2] }{t (t-M_1^2)(t-M_2^2)} \,.
  \nonumber
\end{eqnarray}
It is then seen that, by neglecting the second term suppressed by
$m_{\rm eff}^\nu$, the hierarchical mass pattern 
$M_1 \ll M_2$ enhances the matrix element of the process.
In this limit, $A_t$ 
is dominated by the contribution of $N_1$ as
\begin{eqnarray}
A_t \simeq \frac{ M_1 \, |\Theta_e|^2}{t - M_1^2},
\end{eqnarray}
which leads to the cross section
\begin{eqnarray}
  \label{eq:UB_SIGap}
  \sigma_2 \simeq \frac{G_F^2 \, s \, |\Theta_e|^4}{8 \pi}
   H (r_1)\,,
\end{eqnarray}
where $r_1 = M_1^2/s$ and $H (r_1)$ 
is
$H \simeq 6r_1$ for $M_1 \ll m_W$, 
$H \simeq 2r_1$ for $m_W \ll M_1 \ll \sqrt{s}$, 
and $H \simeq 1/(2r_1)$ for $M_1 \gg \sqrt{s}$.
By taking $m_W/\sqrt{s} \to 0$, it is approximately given by
\begin{eqnarray}
  H (r) =  r(2 + 3 r)
  \left[ 
    \frac{1}{1+r} - \frac{2 r }{ 1+2r}
    \ln \left(1+\frac{1}{r} \right)
  \right] \,.
\end{eqnarray}
It is then found that $H(r_1)$ takes its maximal value 
\begin{eqnarray}
   H (r_1) |_{\rm MAX} = 0.206 
\end{eqnarray}
at $r_1 = 0.542$ ({\it i.e.}, $M_1 = 0.736 \sqrt{s}$).
If we applied the upper bound on $|\Theta_e|^2$ given in Eq.~(\ref{eq:THe_EW}),  the maximal cross section 
$\sigma_2 = 17$~fb could be obtained for 
$\sqrt{s}=3$ TeV.  Unfortunately, such a large cross section cannot 
be realized for the case with ${\cal N}=2$, because we have to take
into account the severe constraint from the $0\nu\beta\beta$ decay.

\begin{figure}
\includegraphics[scale=1.4]{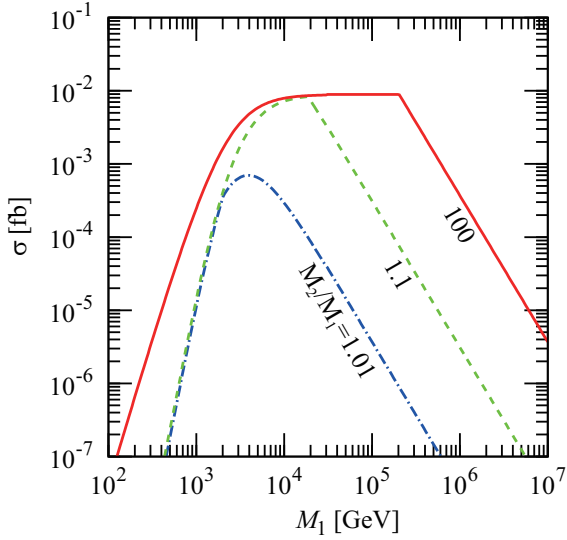}
\caption{\label{fig:SIG_2RHN_wSS_M2dep} 
Cross section of $e^- e^- \to W^- W^-$ in the model with 
two right-handed neutrinos (${\cal N}=2$) for $\sqrt{s} = 3$ TeV.
We take $M_2 = 1.01 M_1$ (the blue dotted-dashed line),
$M_2 = 1.1 M_2$ (the green dashed line) and 
$M_2 = 100 M_2$ (the red solid line). 
Here we consider the inverted hierarchy case 
with $m_{\rm eff}^{\rm UB} = 0.276$~eV
and $|m_{\rm eff}^\nu|$ = $4.79 \times 10^{-3}$~eV.
}
\end{figure}

The effective neutrino mass in this case is estimated as
\begin{eqnarray}
  m_{\rm eff} &=&
  (f_1 - f_2 ) \frac{M_1  \, M_2}{ M_1 + M_2 } |\Theta_e|^2
  \\
  &&
  + 
  \left( 1 - \frac{f_1 \, M_1 + f_2 M_2}{M_1+M_2} \right)
  m_{\rm eff}^\nu \, \,.
  \nonumber 
\end{eqnarray}
The upper bound on $|\Theta_e|^2$ is then given by
Eq.~(\ref{eq:THe_EW}) for $M_1 > M_\ast$, while the more stringent bound 
from the $0\nu\beta\beta$ decay%
\footnote{
When the masses of $N_1$ and $N_2$ are sufficiently degenerate,
we can avoid the constraint from the $0\nu\beta\beta$ decay.
In this case, however,
$A_t$ in Eq.~(\ref{eq:2RHN_At}) approaches to 
(\ref{eq:1RHN_at}), which leads to 
the suppression of the cross section
(see also Fig.~\ref{fig:SIG_2RHN_wSS_M2dep}).
}
\begin{eqnarray}
  |\Theta_e|^2
  <
  \frac{(m_{\rm eff}^{\rm UB} +| m_{\rm eff}^\nu|) \, M_1 \, M_2 }
  { \langle p^2 \rangle \, (M_2 - M_1)} \,,
\end{eqnarray}
is imposed for $M_1 < M_\ast$
where $M_\ast$ is determined from the ratio $M_1/M_2$ as
\begin{eqnarray}
  M_\ast =
  \left( 1 - \frac{M_1}{M_2} \right) \, 
  \frac{ |\Theta_e|^2_{\rm EW} \, \langle p^2 \rangle }
  { m_{\rm eff}^{\rm UB} +| m_{\rm eff}^\nu| } \,.
\end{eqnarray}
Here we have assumed that $M_1^2, M_2^2 \gg \langle p^2 \rangle$.

By using the upper bound on the mixing angle listed above, we can estimate
the maximal value of the cross section of $e^- e^- \to W^- W^-$.
In Fig.~\ref{fig:SIG_2RHN_wSS_M2dep} 
we show how this maximal value depends
on the mass ratio when $\sqrt{s} = 3$ TeV. 
Here we choose $M_2/M_1 = 1.01$, 1.1 and 100.
It can be seen that the largest cross section is obtained 
by taking $M_2/M_1 \to \infty$. (Note that
the difference in the cross section between
$M_2/M_1 = \infty$ and 100 is very small.)

\begin{figure}
\includegraphics[scale=1.4]{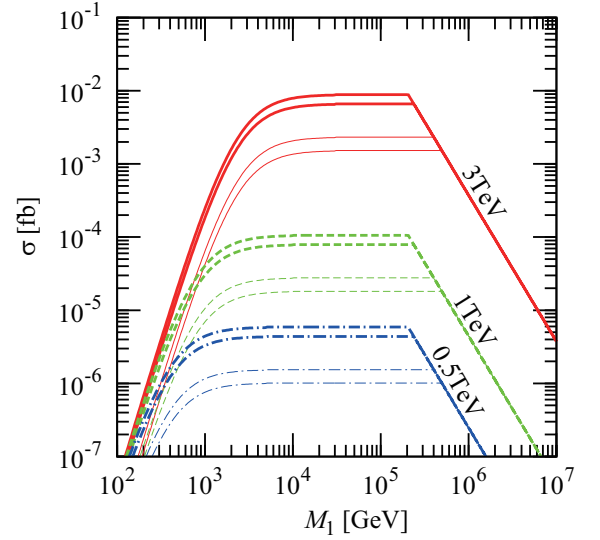}
\caption{\label{fig:SIG_2RHN_wSS_UB} 
Cross section of $e^- e^- \to W^- W^-$ in the model with 
two right-handed neutrinos with $M_1 \ll M_2$
for $\sqrt{s} = 3$ TeV (the red solid lines),
1 TeV (the green dashed lines), and 0.5 TeV (the blue dotted-dashed lines).
In each set of lines, the upper one is for the inverted hierarchy
case with $|m_{\rm eff}^\nu|= 4.79 \times 10^{-3}$~eV
and the lower one is for the normal hierarchy case
with $|m_{\rm eff}^\nu|= 3.66 \times 10^{-3}$~eV.
We use $m_{\rm eff}^{\rm UB}$ = 0.276 and 0.185~eV 
for the thick and thin lines.
}
\end{figure}
We then show in Fig.~\ref{fig:SIG_2RHN_wSS_UB} 
the upper bound on the cross section by taking
$\sqrt{s} = 0.5$, 1 and 3 TeV.
Notice that, when $M_1 < M_\ast= {\cal O} (10^5)$~GeV, the maximal cross section is given by
\begin{align}
  \sigma_2 
  = \frac{G_F^2 \, (m_{\rm eff}^{\rm UB} +| m_{\rm eff}^\nu|)^2}{8 \pi}
  \frac{s^2}{\langle p^2 \rangle^2} \, r_1\, H(r_1) \,.
\end{align}
The function $r_1 \, H(r_1)$ becomes constant for $r_1 \gg 1$.
For $M_1 > M_\ast$ the cross section falls as 
$r_1^{-1}$.
The maximal cross section is then given by
\begin{eqnarray}
  \label{eq:UB_SIG_2RHN}
  \sigma_2 &<&
  8.9 \times 10^{-3} \mbox{fb}
  \left( 
    \frac{m_{\rm eff}^{\rm UB} + |m_{\rm eff}^\nu|}
    {(0.276+0.048)~\mbox{eV}}
  \right)^2
  \nonumber \\
  &&
  \times
  \left(
    \frac{0.178~\mbox{GeV}}{\sqrt{\langle p^2 \rangle}}
  \right)^4
  \left(
    \frac{\sqrt{s}}{3~\mbox{TeV}}
  \right)^4 \,.
\end{eqnarray}

The maximal value changes depending on the mass
hierarchy of active neutrinos.  The cross section in the inverted
hierarchy can be larger than that in the normal hierarchy since
$|m_{\rm eff}^\nu|$ can be larger: $|m_{\rm eff}^\nu| < 3.66 \times
10^{-3}$ eV for the normal hierarchy and $|m_{\rm eff}^\nu| < 4.79
\times 10^{-2}$ eV for the inverted hierarchy.  Here we have estimated
these values by using the central values of the mass squared
differences and the mixing angles of active neutrinos
in Ref.~\cite{Gonzalez-Garcia:2014bfa} and by varying the possible values
of the $CP$ violating phases in the PMNS matrix.  Notice that, as shown
in Fig.~\ref{fig:SIG_2RHN_wSS_UB}, there exists an uncertainty in the
upper bound on the cross section from the nuclear matrix element 
shown in Eq.~(\ref{eq:meffUB}).

The upper bound on the cross section scales as $s^2$ 
and hence the collisions of electrons with higher energies 
are desired.  In fact, the possible way to observe
the inverse $0\nu\beta\beta$ decay is to realize $\sqrt{s}=3$ TeV
at CLIC with the integrated luminosity larger than ${\cal O}(10^2)$ fb.
In such a case, the number of the signal event  can be larger than unity.

\subsection{Case with three right-handed neutrinos}
Finally, let us discuss the case with three right-handed neutrinos
(${\cal N}=3$).  We will show that this case allows the larger
cross section than the previous cases. 
Our basic idea is the following: Let consider three HNLs with
hierarchical masses $M_1 \ll M_2 \ll M_3$.  Then, the large cross
section of $e^- e^- \to W^- W^-$ can be induced by $N_2$ with mass
$M_2 \sim \sqrt{s}$ and mixing $|\Theta_{e2}|^2 \simeq
|\Theta_{e}|^2_{\rm EW}$ (\ref{eq:THe_EW}).  The seesaw relation can be realized
even with the large mixing $|\Theta_{e2}|^2$ because of the
cancellation between $N_2$ and $N_3$, which is
similar to the case with ${\cal N}=2$.  In addition, the constraint
from the $0\nu\beta\beta$ decay can be avoided by the cancellation in
$m_{\rm eff}$ between $N_2$ and $N_1$.  This is
the reason why ${\cal N} \ge 3$ is required for having the large cross
section of the inverse $0\nu\beta\beta$ decay in the framework of the
seesaw mechanism.

Now, we discuss this point in detail.   By using the seesaw relation (\ref{Eseesaw2})
and the effective mass (\ref{eq:meff}), we may express the mixing angles
of $N_1$ and $N_3$ in terms of that of $N_2$:
\begin{eqnarray}
  \Theta_{e1}^2 &=&
  \frac{- 1}{(f_1 -f_3) \, M_1}
  \Bigl[ (f_2 - f_3) \, M_2 \, \Theta_{e2}^2
  \nonumber \\
  && ~~~~~~~~~ - m_{\rm eff} + ( 1 - f_3) \, m_{\rm eff}^\nu   \Bigr] \,,
  \\
  \Theta_{e3}^2 &=&
  \frac{-1}{(f_1 -f_3) \, M_3}
  \Bigl[ (f_1 - f_2) \, M_2 \, \Theta_{e2}^2 
  \nonumber \\
  && ~~~~~~~~~   + m_{\rm eff} - ( 1 - f_1) \, m_{\rm eff}^\nu \Bigr] \,,
\end{eqnarray}
It is then found that $A_t$ is
\begin{eqnarray}
  \label{eq:At_3RHN}
  A_t &=&
  - 
  \frac{ M_2 \, (M_2^2 - M_1^2) \, (M_3^2 - M_2^2) \,
    (t + \langle p^2 \rangle) \, \Theta_{e2}^2}
  { (M_2^2 + \langle p^2 \rangle ) \,
    (t - M_1^2) \, (t - M_2^2) \, (t - M_3^2) }
  \nonumber \\  
  &&
  - \frac{ m_{\rm eff} \, ( M_1^2 + \langle p^2 \rangle )
  \, ( M_3^2 + \langle p^2 \rangle )}
  {\langle p^2 \rangle \, (t - M_1^2) \, (t - M_3^2)} 
  \nonumber \\
  &&
  + \frac{ m_{\rm eff}^\nu \, M_1^2 M_3^2 \, (t + \langle p^2 \rangle)}
  { \langle p^2 \rangle \, t \, ( t- M_1^2 )\, (t -M_3^2)} \,.
\end{eqnarray}
By neglecting the second and third terms suppressed by
$m_{\rm eff}$ and $m_{\rm eff}^\nu$, 
the matrix element can be maximal when the masses of three HNLs
are hierarchical, namely $M_1 \ll M_2 \ll M_3$.
In this case the mixing angles become
\begin{eqnarray}
  \Theta_{e1}^2 &\simeq& - \frac{M_1}{M_2} \Theta_{e2}^2 \,,
  \\
  \Theta_{e3}^2 &\simeq& - \frac{M_2}{M_3} \Theta_{e2}^2 \,,
\end{eqnarray}
and hence $|\Theta_{e1}|^2$, $|\Theta_{e3}|^2 \ll |\Theta_{e2}|^2$, 
so $|\Theta_{e2}|^2 \simeq |\Theta_{e}|^2$,
which results in 
\begin{eqnarray}
  A_t \simeq \frac{M_2 \, |\Theta_{e}|^2}{t - M_2^2}  \,,
\end{eqnarray}
In this limit the cross section is given by
\begin{eqnarray}
  \label{eq:SIG_3RHN}
  \sigma_3  \simeq \frac{G_F^2 \, s \, |\Theta_e|^4}{8 \pi}
   H (r_2) \,,
\end{eqnarray}
where $r_2 = M_2^2/s$. The result is similar to
Eq.~(\ref{eq:UB_SIGap}) for ${\cal N}=2$.

\begin{figure}
\includegraphics[scale=1.4]{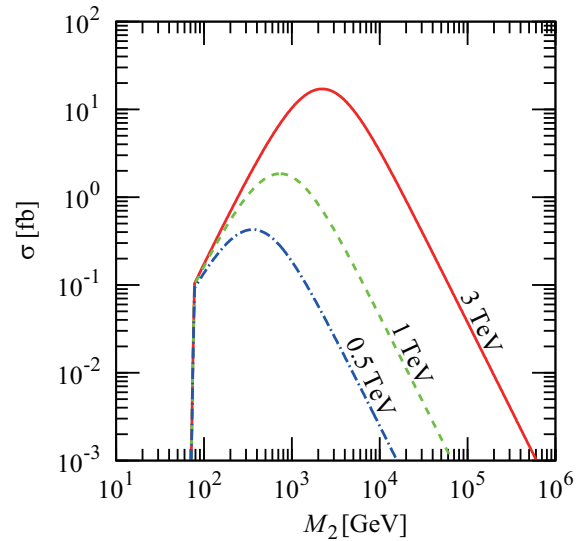}
\caption{\label{fig:SIG_3RHN_wSSwLEP_UB_v3} 
Cross section of $e^- e^- \to W^- W^-$ in the model with 
three right-handed neutrinos (${\cal N} = 3$) with $M_1 \ll M_2 \ll M_3$
for $\sqrt{s} = 3$ TeV (the red solid line),
1 TeV (the green dashed line), and 0.5 TeV (the blue dotted-dashed line).
}
\end{figure}
Thus, the upper bound on the cross section can be obtained 
by the maximal value of the mixing angle,
which is shown in Fig.~\ref{fig:SIG_3RHN_wSSwLEP_UB_v3}.
Here we take $M_1 = 3$ GeV to avoid the stringent constraint
on $|\Theta_{e1}|^2$ from the direct search experiments.
Notice that the bound on $|\Theta_e|^2$ comes from 
the search at LEP~\cite{Abreu:1996pa} for $M_2 \lesssim m_Z$, 
and comes from the electroweak precision test (\ref{eq:THe_EW}) 
for $M_2 \gtrsim m_Z$.  Since the former bound is so stringent, the cross section is too small to be observed for that mass range.
The maximal cross section is obtained when
$M_2 \simeq 0.736 \sqrt{s}$ as
\begin{eqnarray}
  \label{eq:sig_3}
  \sigma_3 <
  17 \, \mbox{fb} \,
  \left( \frac{\sqrt{s}}{3 \, \mbox{TeV}} \right)^2
  \left( \frac{|\Theta_{e}|^2}{|\Theta_{e}|^2_{\rm EW}} \right)^2 \,.
\end{eqnarray}
Compared with Eq. (\ref{eq:UB_SIG_2RHN}), the maximal value of the cross
section in the ${\cal N}=3$ case can be much larger than the previous
cases.  

\begin{figure}
\includegraphics[scale=1.4]{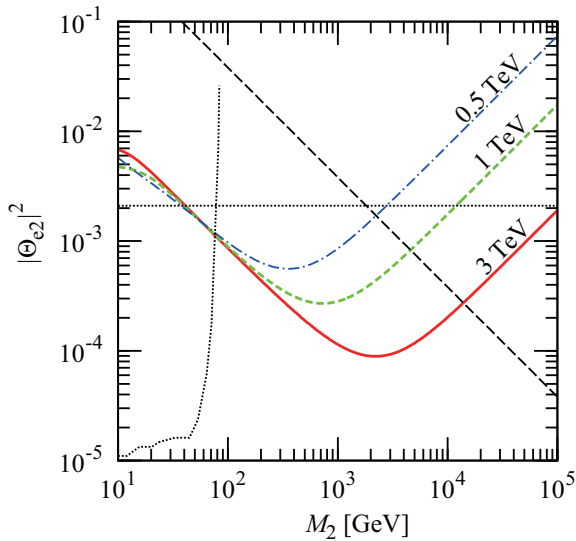}
\caption{\label{fig:SIG_3RHN_wSSwLEP_UBTHesq_v3} 
Sensitivity limit of $|\Theta_{e2}|^2$ 
from $e^- e^- \to W^- W^-$ in the model with 
three right-handed neutrinos (${\cal N} = 3$) 
for $\sqrt{s} = 3$ TeV (the red solid line),
1 TeV (the green dashed line), and 0.5 TeV (the blue dotted-dashed line).
The black dotted lines show the upper bounds
from Refs.~\cite{Abreu:1996pa} and \cite{Antusch:2014woa}. The black long-dashed line shows the upper bound
from the perturbativity of Yukawa couplings when
$M_3 = 10^5$~GeV.
}
\end{figure}
In Fig.~\ref{fig:SIG_3RHN_wSSwLEP_UBTHesq_v3}, we show the sensitivity
limit of $|\Theta_{e2}|^2$ together with the current upper limits (for the perturbativity bound $|\Theta_{e2}|^2<4\pi\langle\Phi\rangle^2/(M_2M_3)$, see Ref. \cite{Asaka}).  
Interestingly, the
inverse $0\nu\beta\beta$ decay process can probe HNL ($N_2$ in this
case) of which the mass is much heavier than the center of mass energy ($M_2
\gg \sqrt{s}$), because it is induced as the virtual effect of HNLs.
This is one of the most important advantages of $e^- e^- \to W^- W^-$
in searching HNLs~\cite{Belanger:1995nh,Greub:1996ct}.

It should be noted that the process (\ref{eeeww})
is induced by the collision of two electrons with negative helicity.
Thus, this reaction will be turned off simply by using 
one of the electron beams with positive helicity~\cite{London:1987nz}.
This can offer the important test for this process
by using the polarized beam which is specific to the linear
colliders.

We should note that the large cross section shown in Eq. (\ref{eq:sig_3})
is possible when the HNL $N_1$ with $M_1 \ll M_2$ and
$\Theta_{e1}^2 \simeq - (M_1/M_2)\, \Theta_{e2}^2$ exists in order to
avoid the stringent constraint from the $0\nu\beta\beta$ decay.  If this
is the case, such a light HNL $N_1$ is also a good target for the
future experiments.  The searches by using charmed meson
decay~\cite{Alekhin:2015byh,Anelli:2015pba}, $e^+ e^- \to \nu
N_1$~\cite{Buchmuller:1991tu,Banerjee:2015gca} and $Z \to \nu
N_1$~\cite{Blondel:2014bra} may provide the complementary test for the
model discussed in this analysis.

Here, we have derived the sensitivity limit of the mixing by
just counting the numbers of the signal event assuming no
background event.  The possible background processes have been studied
in~\cite{London:1987nz,Greub:1996ct,Banerjee:2015gca}, which may
reduce the sensitivity.  However, the usage of the polarized beam can
make the cross section four times larger and the longer duration of
the experiment can increase the number of events.  These issues must be
studied in detail to have the precise sensitivity limit; however, they
are beyond the scope of this analysis.

So far, we have discussed the case with the hierarchical mass pattern
$M_1 \ll M_2 \ll M_3$ with $M_2 \sim \sqrt{s}$.  When $M_2 \gg
\sqrt{s}$, $N_2$ decouples from the process and the cross section
becomes small.  Even in this case,
when $M_1 \sim \sqrt{s} \ll M_2 \ll M_3$,
the situation becomes the same as the case with ${\cal N} = 2$.
Then, the maximal cross section is given by Eq.~(\ref{eq:UB_SIG_2RHN}).

Note that Eq.~(\ref{eq:SIG_3RHN}) is independent from
$m_{\rm eff}$.  This means that the inverse $0\nu\beta\beta$ decay can
happen even if the
the $0\nu\beta\beta$ decay could not be observed.  This is
because the energy scales of the former process ($\sqrt{s} \sim 1$
TeV) and the latter one ($\sqrt{\langle p^2 \rangle} \sim 0.1$ GeV)
are much different.  Therefore, the $0\nu\beta\beta$ decay and the
inverse one are regarded as complementary tests for
the violation of lepton number in nature.

Finally, we comment on the case with ${\cal N} >3$.
The maximal cross section is the same as that with ${\cal N}=3$.
This is because the severe constraints from the seesaw relation
and the $0\nu\beta\beta$ decay can be avoided as described above, 
but the inevitable upper bound (\ref{eq:THe_EW}) on 
$|\Theta_e|^2$, which is just the sum of $|\Theta_{eI}|^2$,
gives the cross section as shown in Eq.~(\ref{eq:sig_3}).

\section{Conclusions} \label{sCon} 
We have discussed the inverse $0\nu\beta\beta$ decay process
$e^- e^- \to W^- W^-$ in the SM with $\mathcal{N}$ right-handed
neutrinos.  We have estimated for the first time the maximal cross sections 
of this process depending on $\mathcal{N}$ by 
taking into account the seesaw relation on the mixings of 
active neutrinos and HNLs as well as the severe constraints 
from direct search experiments, electroweak precision measurements
and $0\nu\beta\beta$ decay. The maximal cross sections are realized in limited regions of the
parameter space, and  the cross sections are orders of magnitude smaller if there is no fine-tuning.

For the $\mathcal{N}=1$ case, the seesaw relation
restricts the cross section to be smaller than ${\cal O} (10^{-18})$~fb, so
it cannot be seen at future experiments. For the $\mathcal{N}=2$ case,
the cross section can be 
$\sigma_{2} \sim 10^{-2}$ fb for $\sqrt{s}=3$~TeV. It may then be
observable at CLIC, but it is impossible at ILC. For the
$\mathcal{N}=3$ case, the larger cross section as
$\sigma_{3}=$ 0.47 (17) fb for
$\sqrt{s}=$ 0.5 (3) TeV can be obtained.  In this case
both ILC and CLIC can search the inverse
$0\nu\beta\beta$ decay,
which would reveal the fate of the lepton number
and the origin of the neutrino masses.

\section*{ACKNOWLEDGEMENTS}
T.A. was supported by JSPS KAKENHI Grants No.
25400249, No. 26105508, and No. 15H01031.

\appendix
\section{DEFINITION OF VARIABLES}
In this Appendix we define the variables used in the cross section of
$e^- e^- \to W^- W^-$ in Eq. (\ref{Edsig}).  Here, we work in the
center-of-mass frame and $\theta$ is a scattering angle of one of the
outgoing $W^-$s.  Mandelstam variables are defined as usual,
\begin{align}
t&=-\frac{s}{2}(1-\beta_W\cos\theta)+m_W^2,\\
u&=-\frac{s}{2}(1+\beta_W\cos\theta)+m_W^2,
\end{align}
where $\beta_W \equiv \sqrt{1-4r_W}$ with
$r_W\equiv m_W^2/s$.
The variables $A_t$ and $A_u$ are given by
\begin{align}
  A_t &= \sum_{i = 1}^3 \frac{U_{ei}^2 \, m_i}{t - m_i^2} 
  + \sum_{I=1}^{\cal N} \frac{\Theta_{eI}^2 \, M_I}{t - M_I^2}  \,,
  \\
  A_u &= \left. A_t \right|_{ t \to u} \,.
\end{align}
We have found a difference in the definition of $A_t$
from Refs.~\cite{Gluza:1995ix,Gluza:1995js}.
Here we have corrected the errors in the denominators of $A_t$
as $2 t \to t$.

The range of $|t|$ is given by $ (1 - 2 \, r_W + \beta_W) s /2 \ge |t|
\ge (1 - 2 \, r_W - \beta_W) s /2$ and then $|t| \ge m_W^2 \, [r_W +
{\cal O}(r_W^2)]$.  Therefore, $m_i^2/|t| \ll 1$ for the realistic
range of $\sqrt{s}$.  In this case, $A_t$ is 
\begin{eqnarray}
  A_t 
  =
  \frac{m_{\rm eff}^\nu}{t} 
  + \sum_{I=1}^{\cal N} \frac{\Theta_{eI}^2 \, M_I}{t - M_I^2} \,.
\end{eqnarray}
On the other hand, the variables $B_t$, $B_u$ and $B_{tu}$ are
given by
\begin{eqnarray}
  \label{EBt}
  B_t &=& ( 1 - 4 \, r_W ) \, t^2 - 4 \, r_W ( 1- 2 \, r_W) \, s\, t
  \nonumber \\
  &&  + \, 4 (  1 - r_W ) \, r_W^2 \, s^2 \,,
  \\
  B_u &=& \left. B_t \right|_{t \to u} \,,
  \\
  B_{tu} &=& ( 1 - 4 \, r_W) \, t \, u + 4 \, r_W^3 \, s^2 \,.
\end{eqnarray}
A typo in Ref. \cite{Rodejohann:2010jh} is corrected in Eq. (\ref{EBt}).

Since we are considering $\sqrt{s}\geq 500$ GeV, we can approximate
$m_W^2\ll s$. We cannot, however, approximate $m_W^2\ll |t|$, since
$|t|\simeq m_W^4/s<m_W^2$ at $\theta\simeq 0$. Indeed,
\begin{align}
  \int_{-1}^1 d\cos\theta\frac{m_W^4}{t^2}
  =
  2 + {\cal O}(r_W) \,,
\end{align}
so we have to take the limit $m_W^2/s\to 0$ after integration when we
consider the contribution of light neutrinos. This is the reason that
the cross section in Eq. (\ref{eq:sig_0}) is three times larger than 
that in previous
studies \cite{Belanger:1995nh,Rodejohann:2010jh}.




\begin{thebibliography}{100}{

\bibitem{Gonzalez-Garcia:2014bfa} 
  M.~C.~Gonzalez-Garcia, M.~Maltoni and T.~Schwetz,
  JHEP {\bf 1411}, 052 (2014)
  [arXiv:1409.5439 [hep-ph]].

\bibitem{Minkowski:1977sc} 
  P.~Minkowski,
  Phys.\ Lett.\ B {\bf 67}, 421 (1977).

\bibitem{Yanagida:1979as} 
T. Yanagida, in {\it Proceedings of the Workshop on Unified Theory and Baryon Number of the Universe}, edited by O. Sawada and A. Sugamoto (KEK, Tsukuba, Ibaraki 305-0801, 1979), p. 95.

\bibitem{Yanagida:1980xy} 
  T.~Yanagida,
  Prog.\ Theor.\ Phys.\  {\bf 64}, 1103 (1980).

\bibitem{GellMann:1980vs} 
  M. Gell-Mann, P. Ramond, and R. Slansky, in {\it Supergravity}, edited by P. van Niewenhuizen and D. Freedman (North Holland, Amsterdam, 1979)
  [arXiv:1306.4669 [hep-th]].
  
  

\bibitem{Ramond:1979}
P.~Ramond, 
in  Talk given at the Sanibel Symposium, 
Palm Coast, Fla., Feb.~25-Mar.~2, 1979, preprint CALT-68-709
(retroprinted as hep-ph/9809459).



\bibitem{Glashow:1979nm} 
  S.~L.~Glashow,
  NATO Sci.\ Ser.\ B {\bf 61}, 687 (1980).

\bibitem{Mohapatra:1979ia} 
  R.~N.~Mohapatra and G.~Senjanovic,
  Phys.\ Rev.\ Lett.\  {\bf 44}, 912 (1980).

\bibitem{Fukugita:1986hr}
  M.~Fukugita and T.~Yanagida,
  Phys.\ Lett.\  B {\bf 174} (1986) 45 .

\bibitem{Giudice:2003jh}
  For example, see
  G.~F.~Giudice, A.~Notari, M.~Raidal, A.~Riotto and A.~Strumia,
  Nucl.\ Phys.\  B {\bf 685} (2004) 89
  [arXiv:hep-ph/0310123].

\bibitem{Leptogenesis_inflation_decay}
  G.~Lazarides and Q.~Shafi,
  Phys.\ Lett.\ B {\bf 258} (1991) 305;


 
  T.~Asaka, K.~Hamaguchi, M.~Kawasaki and T.~Yanagida,
  Phys.\ Lett.\ B {\bf 464} (1999) 12
  [hep-ph/9906366];
  Phys.\ Rev.\ D {\bf 61} (2000) 083512
  [hep-ph/9907559].


\bibitem{Pilaftsis:2003gt} 
  A.~Pilaftsis and T.~E.~J.~Underwood,
  Nucl.\ Phys.\ B {\bf 692}, 303 (2004)
  [hep-ph/0309342].

\bibitem{Akhmedov:1998qx}
  E.~K.~Akhmedov, V.~A.~Rubakov and A.~Y.~Smirnov,
  Phys.\ Rev.\ Lett.\  {\bf 81} (1998) 1359.
  

\bibitem{Asaka:2005pn} 
  T.~Asaka and M.~Shaposhnikov,
  Phys.\ Lett.\ B {\bf 620}, 17 (2005)
  [hep-ph/0505013].

\bibitem{Asaka:2013jfa} 
  T.~Asaka and S.~Eijima,
  PTEP {\bf 2013}, no. 11, 113B02 (2013)
  [arXiv:1308.3550 [hep-ph]].
  
\bibitem{Dodelson:1993je} 
  S.~Dodelson and L.~M.~Widrow,
  Phys.\ Rev.\ Lett.\  {\bf 72}, 17 (1994)
  [hep-ph/9303287].

\bibitem{Pulsarkick}
  A.~Kusenko and G.~Segre,
  Phys.\ Lett.\  B {\bf 396} (1997) 197
  [arXiv:hep-ph/9701311];
  G.~M.~Fuller, A.~Kusenko, I.~Mocioiu and S.~Pascoli,
  Phys.\ Rev.\  D {\bf 68} (2003) 103002
  [arXiv:astro-ph/0307267].


\bibitem{Kusenko:2009up}
  A.~Kusenko,
  Phys.\ Rept.\  {\bf 481} (2009) 1
  [arXiv:0906.2968 [hep-ph]].


\bibitem{Fuller:2009zz} 
  G.~M.~Fuller, A.~Kusenko and K.~Petraki,
  Phys.\ Lett.\ B {\bf 670}, 281 (2009)
  [arXiv:0806.4273 [astro-ph]].  

\bibitem{Georgi}
H. Georgi, in {\it Particles and Fields}, edited by C. E. Carlson, AIP Conf. Proc. No. 23 (AIP, New York, 1975)  
\bibitem{Fritzsch:1974nn} 
  H.~Fritzsch and P.~Minkowski,
  Annals Phys.\  {\bf 93}, 193 (1975).

  
\bibitem{Tsuyuki:2014xja} 
  T.~Tsuyuki,
  PTEP {\bf 2015}, 011B01 (2015)
  [arXiv:1411.2769 [hep-ph]].
  
\bibitem{Clarke:2015gwa} 
  J.~D.~Clarke, R.~Foot and R.~R.~Volkas,
  Phys.\ Rev.\ D {\bf 91}, no. 7, 073009 (2015)
  [arXiv:1502.01352 [hep-ph]].

\bibitem{Kusenko:2004qc}
  A.~Kusenko, S.~Pascoli and D.~Semikoz,
  JHEP {\bf 0511} (2005) 028
  [arXiv:hep-ph/0405198].

\bibitem{Gorbunov:2007ak}
  D.~Gorbunov and M.~Shaposhnikov,
  JHEP {\bf 0710} (2007) 015
  [arXiv:0705.1729 [hep-ph]].

\bibitem{Atre:2009rg} 
  A.~Atre, T.~Han, S.~Pascoli and B.~Zhang,
  JHEP {\bf 0905}, 030 (2009)
  [arXiv:0901.3589 [hep-ph]].

\bibitem{Deppisch:2015qwa} 
  F.~F.~Deppisch, P.~S.~Bhupal Dev and A.~Pilaftsis,
  New J.\ Phys.\  {\bf 17}, no. 7, 075019 (2015)
  [arXiv:1502.06541 [hep-ph]].


\bibitem{Pas:2015eia} 
  See, for example, a recent review,
  H.~Pas and W.~Rodejohann,
  arXiv:1507.00170 [hep-ph].

\bibitem{Rizzo:1982kn} 
  T.~G.~Rizzo,
  Phys.\ Lett.\ B {\bf 116}, 23 (1982).


\bibitem{Baer:2013cma} 
  H.~Baer {\it et al.},
  arXiv:1306.6352 [hep-ph].
  
\bibitem{Accomando:2004sz} 
  E.~Accomando {\it et al.} [CLIC Physics Working Group Collaboration],
  hep-ph/0412251.



\bibitem{London:1987nz} 
  D.~London, G.~Belanger and J.~N.~Ng,
  Phys.\ Lett.\ B {\bf 188}, 155 (1987).

\bibitem{Dicus:1991fk} 
  D.~A.~Dicus, D.~D.~Karatas and P.~Roy,
  Phys.\ Rev.\ D {\bf 44}, 2033 (1991).

\bibitem{Belanger:1995nh}
  G.~Belanger, F.~Boudjema, D.~London and H.~Nadeau,
  Phys.\ Rev.\ D {\bf 53}, 6292 (1996) 
  [hep-ph/9508317].

\bibitem{Gluza:1995ky} 
  J.~Gluza and M.~Zralek,
  Phys.\ Rev.\ D {\bf 52}, 6238 (1995)
  [hep-ph/9502284].

\bibitem{Gluza:1995ix} 
  J.~Gluza and M.~Zralek,
  Phys.\ Lett.\ B {\bf 362}, 148 (1995)
  [hep-ph/9507269].

\bibitem{Gluza:1995js} 
  J.~Gluza and M.~Zralek,
  Phys.\ Lett.\ B {\bf 372}, 259 (1996)
  [hep-ph/9510407].

  
\bibitem{Greub:1996ct} 
  C.~Greub and P.~Minkowski,
  eConf C {\bf 960625}, NEW149 (1996)
  [Int.\ J.\ Mod.\ Phys.\ A {\bf 13}, 2363 (1998)]
  [hep-ph/9612340].

\bibitem{Rodejohann:2010jh}
  W.~Rodejohann,
  Phys.\ Rev.\ D {\bf 81} (2010) 114001
  [arXiv:1005.2854 [hep-ph]].
  
\bibitem{Banerjee:2015gca} 
  S.~Banerjee, P.~S.~B.~Dev, A.~Ibarra, T.~Mandal and M.~Mitra,
  arXiv:1503.05491 [hep-ph].
  
\bibitem{Grimus:2002nk} 
  W.~Grimus and L.~Lavoura,
  Phys.\ Lett.\ B {\bf 546}, 86 (2002)
  [hep-ph/0207229].
  
\bibitem{AristizabalSierra:2011mn} 
  D.~Aristizabal Sierra and C.~E.~Yaguna,
  JHEP {\bf 1108}, 013 (2011)
  [arXiv:1106.3587 [hep-ph]].
  
\bibitem{Faessler:2014kka} 
  A.~Faessler, M.~Gonzalez, S.~Kovalenko and F.~\v Simkovic,
  Phys.\ Rev.\ D {\bf 90}, no. 9, 096010 (2014)
  [arXiv:1408.6077 [hep-ph]].


\bibitem{Gando:2012zm} 
  A.~Gando {\it et al.} [KamLAND-Zen Collaboration],
  Phys.\ Rev.\ Lett.\  {\bf 110}, no. 6, 062502 (2013)
  [arXiv:1211.3863 [hep-ex]].


\bibitem{Agostini:2013mzu} 
  M.~Agostini {\it et al.}  [GERDA Collaboration],
  Phys.\ Rev.\ Lett.\  {\bf 111}, no. 12, 122503 (2013)
  [arXiv:1307.4720 [nucl-ex]].



%
\bibitem{Abreu:1996pa} 
  P.~Abreu {\it et al.} [DELPHI Collaboration],
  Z.\ Phys.\ C {\bf 74}, 57 (1997)
  [Z.\ Phys.\ C {\bf 75}, 580 (1997)].

\bibitem{Antusch:2014woa}
  S.~Antusch and O.~Ficher,
  JHEP {\bf 1410} (2014) 94
  [arXiv:1407.6607 [hep-ph]].

\bibitem{Adam:2013mnn} 
  J.~Adam {\it et al.} [MEG Collaboration],
  Phys.\ Rev.\ Lett.\  {\bf 110}, 201801 (2013)
  [arXiv:1303.0754 [hep-ex]].

\bibitem{Agashe:2014kda} 
  K.~A.~Olive {\it et al.} [Particle Data Group Collaboration],
  Chin.\ Phys.\ C {\bf 38}, 090001 (2014).

\bibitem{Gluza:2002vs} 
  J.~Gluza,
  Acta Phys.\ Polon.\ B {\bf 33}, 1735 (2002)
  [hep-ph/0201002].

\bibitem{Kersten:2007vk} 
  J.~Kersten and A.~Y.~Smirnov,
  Phys.\ Rev.\ D {\bf 76}, 073005 (2007)
  [arXiv:0705.3221 [hep-ph]].

\bibitem{Asaka} 
  T.~Asaka and T.~Tsuyuki,
  arXiv:1509.02678 [hep-ph].
  
\bibitem{Alekhin:2015byh} 
  S.~Alekhin {\it et al.},
  arXiv:1504.04855 [hep-ph].
\bibitem{Anelli:2015pba} 
  M.~Anelli {\it et al.} [SHiP Collaboration],
  arXiv:1504.04956 [physics.ins-det].

\bibitem{Buchmuller:1991tu} 
  W.~Buchmuller and C.~Greub,
  Nucl.\ Phys.\ B {\bf 363}, 345 (1991).


\bibitem{Blondel:2014bra} 
  A.~Blondel {\it et al.} [FCC-ee study Team Collaboration],
  arXiv:1411.5230 [hep-ex].


}\end{thebibliography}
\end{document}